# New Results From UKQCD Using the Cray T3D: Measuring Gluonic Observables *

UKQCD Collaboration - presented by Hartmut Wittig [a]

[a]Physics Department, The University, Southampton SO17 1BJ

We report on a set of simulations performed at several values of the lattice spacing on the Cray T3D at Edinburgh. Different methods to extract the lattice scale from the static quark potential are discussed.

Over the past three years, the UKQCD Collaboration has produced a great many of results in lattice QCD using data that were generated on the 64-node, i860-based Meiko Computing Surface ("Maxwell"). Now the collaboration has embarked on a series of new simulations that are carried out on the recently installed Cray T3D at Edinburgh. This machine is based on the MIMD concept, and currently its version at Edinburgh consists of 128 nodes, each containing two DEC-Alpha processors capable of a peak speed of 300 MFlops and with 128 MBytes of memory. Thus, the peak speed of the whole system is 38 GFlops with a total memory of 16 GBytes. The nodes form a **T**orus in **3 D**imensions with the possibility of bidirectional communication. A fairly unusual feature is the powerful front-end, a Cray Y-MP of which there is a two-processor version at Edinburgh. The Y-MP is linked to 200 GBytes of disk space as well as to a mass data store with a capacity of 1 TByte which is to be upgraded to about 20 TBytes in the future.

In order to ensure portability as well as a long lifetime of the code used in our simulations of quenched lattice QCD, UKQCD has decided to employ standard Fortran77 with PVM for message passing as our programming model [1]. In addition, a High-Performance Fortran version has been developed, which however has not been used in any of the production runs so far.

The current sustained speed is about 4 GFlops for the pure gauge code and about 7 GFlops for the propagator code (measured on the full 256

*Talk presented at *Lattice '94*, Bielefeld, Germany, 27 Sep–1 Oct 1994

Table 1
Lattices simulated, showing the ratio of overrelaxed to Cabibbo-Marinari sweeps, the number of separating sweeps and the number of configurations analysed for this study.

| $\beta$ | $L^3 \cdot T$ | $N_{\rm OR} : N_{\rm CM}$ | $N_{\rm sep}$ | conf. |
|---|---|---|---|---|
| 6.0 | $16^3 \cdot 48$ | 7:1 | 800 | 193 |
| 6.2 | $24^3 \cdot 48$ | 5:1 | 2400 | 101 |
| 6.4 | $32^3 \cdot 64$ | 7:1 | 1000 | 59 |

processor partition). The average I/O speed is 20 MBytes/sec. It should be emphasised that the code is still being tuned towards better performance concerning both speed and I/O.

The availability of a Y-MP as front-end makes it possible to perform some part of the analysis code on the vector processors which takes some load off the main machine. This idea is exploited in the computation of gluonic observables.

The runs on the Cray T3D represent a continuation of UKQCD's earlier simulations, now extended to a range of $\beta$-values and high statistics. A summary of lattice sizes and $\beta$-values is presented in Tab. 1. Our ultimate goal is to study hadron phenomenology including a quantification of residual lattice artefacts using the $O(a)$-improved Sheikholeslami-Wohlert action [2,3].

The gauge configurations were generated on the T3D using the Cabibbo-Marinari heatbath method plus a number of overrelaxation sweeps. They were subsequently analysed on the Y-MP front-end to measure the static quark potential, Polyakov loops and the two lowest glueball states. These measurements mainly serve to set the lattice scale, using either the string tension $\sqrt{K}$ or the hadronic scale $r_0$ whose merits are discussed in [4].

The static quark potential $V(r)$ was extracted by measuring "fuzzed" Wilson loops $W(t,r)$, following the procedure outlined in [4]. At $\beta = 6.0, 6.2$ the orientations (1,0,0), (1,1,0) and (1,1,1) were computed, whereas at $\beta = 6.4$ only (1,0,0) was considered. The fuzzed links were constructed with link/staple ratio of 2:1 and with $L-2$ and $L+8$ iterations of the fuzzing algorithm ($L$ being the spatial lattice size). The two fuzzing levels were taken as a $2 \times 2$ variational basis to extract the eigenvalues and -vectors of the generalised eigenvalue equation

$$W_{ij}(t,r)\, v(r)_j^{(k)} = \lambda_k(t,r)\, W_{ij}(t,r)\, v(r)_j^{(k)}. \quad (1)$$

The eigenvector $v(r)^{(1)}$ corresponding to $\lambda_1(t_0, r)$ at $t_0 = 3$ was used to project onto the approximate ground state. The resulting correlator was then fitted to both single and double exponentials for timeslices up to $t = 8$. Unless stated otherwise the following analysis is based on the results from double exponential fits. Statistical errors are obtained from a bootstrap analysis based on 100 bootstrap samples.

In order to obtain the string tension $\sqrt{K}$, the data for the potential were fitted according to

$$V(r) = V_0 + K|r| - e\, G_L(r), \quad (2)$$

where $G_L(r)$ is the lattice Greens function for one-gluon exchange. Our best estimates for the fitted parameters are shown in Tab. 2. We have included an estimate of systematic errors based on the spread obtained from varying the ground state potential and the minimum distance included in the fit to eq. (2). The value for the string tension are in good agreement with those quoted in refs. [5,6]. In the last column we quote the values obtained from determining the string tension by measuring the correlation of Polyakov loops (i.e. the torelon mass). The signal for the torelon is still relatively poor, but the values are in good agreement with the Wilson loop data and also with ref. [7].

As discussed in [4], an alternative lattice scale is the distance $r_0$ at which

$$F(r_0)\, r_0^2 = 1.65. \quad (3)$$

This relation corresponds to $r_0 \simeq 0.5$ fm in phenomenological non-relativistic potential models describing quarkonia. In order to compute $r_0$, we calculated the tree-level improved force $F_{\vec{d}}(r_I)$ [4] according to

$$F_{\vec{d}}(r_I) = |\vec{d}|^{-1}\big(V(r) - V(r - \vec{d})\big), \quad (4)$$

$$r_I = |\vec{d}|\big(G_L(r) - G_L(r - \vec{d})\big)^{-1}. \quad (5)$$

$r_0$ was then obtained by linearly interpolating the force around $F(r_I)\, r_I^2 = 1.65$ [4]. Tab. 2 shows the results including the estimates of systematic errors from the spread of values obtained using different points in the interpolation and by varying the estimate for the ground state potential.

Our results for $r_0$ and $r_0\sqrt{K}$ are consistent with those quoted in [9] and [8]. However, we note that our statistical errors, in particular at $\beta = 6.4$, are smaller.

Using the values for $r_0$ we can now scale the results at all three $\beta$-values and convert into physical units using $r_0 = 0.5$ fm. This is shown in Fig. 1. The data can be compared to the prediction from the string formula

$$F(r)\, r_0^2 = 1.65 + \frac{\pi}{12}\left(\frac{r_0^2}{r^2} - 1\right) \quad (6)$$

which is derived from (2) using the constraint (3) and $e = \pi/12$. This gives a remarkably good description of the data (solid line). The linearity of the force around 0.5 fm illustrates the fact that $r_0$ can be obtained reliably since one is not required to know the detailed behaviour of the curve at very small or large $|r|$.

The $\beta$-dependence of the results for $r_0$ can be compared to perturbation theory using the two-loop $\beta$-function for the gauge coupling. The value of $r_0$ at $\beta = 6.4$ was used as input, and $\alpha_s$ was taken from ref. [10]. Fig. 2 compares the two-loop prediction with the one-loop formula. It is seen that the two-loop formula describes the data rather well, strengthening our confidence in the consistency of our results.

Using $r_0 = 0.5$ fm and $\sqrt{K} = 440$ MeV we can now give estimates for $a^{-1}$ GeV. The results in Tab. 3 again include our estimates of systematic uncertainties.

Table 2
The fitted parameters for the potential, eq. (2), results for $r_0$, and the string tension from the Polyakov loop ("PL"). The first error is statistical, the second is an estimate of systematic uncertainties.

| $\beta$ | $V_0$ | $K$ | $e$ | $r_0$ | $r_0\sqrt{K}$ | $K_{PL}$ |
|---|---|---|---|---|---|---|
| 6.0 | $0.659\,^{+3}_{-3}\,^{+2}_{-9}$ | $0.0468\,^{+6}_{-6}\,^{+12}_{-4}$ | $0.296\,^{+3}_{-4}\,^{+1}_{-18}$ | $5.53\,^{+7}_{-8}\,^{+1}_{-14}$ | $1.19\,^{+1}_{-2}$ | $0.043\,^{+5}_{-4}$ |
| 6.2 | $0.633\,^{+2}_{-2}\,^{+2}_{-9}$ | $0.0252\,^{+3}_{-3}\,^{+11}_{-4}$ | $0.288\,^{+4}_{-5}\,^{+5}_{-19}$ | $7.33\,^{+17}_{-24}\,^{+0}_{-9}$ | $1.16\,^{+2}_{-3}$ | $0.025\,^{+3}_{-2}$ |
| 6.4 | $0.604\,^{+2}_{-3}\,^{+6}_{-2}$ | $0.0143\,^{+3}_{-3}\,^{+8}_{-1}$ | $0.286\,^{+6}_{-9}\,^{+3}_{-16}$ | $9.70\,^{+11}_{-30}\,^{+5}_{-2}$ | $1.16\,^{+2}_{-3}$ | $0.016\,^{+1}_{-1}$ |

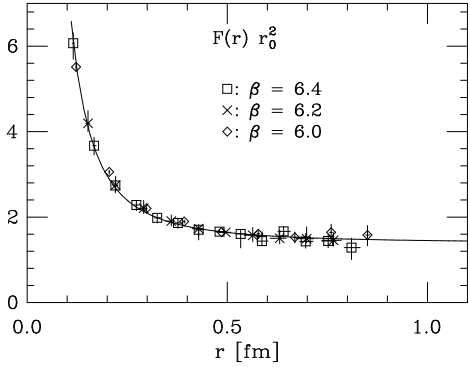

Figure 1. The force at all $\beta$-values, obtained from scaling the results using $r_0$. Statistical errors on $r_0$ are included.

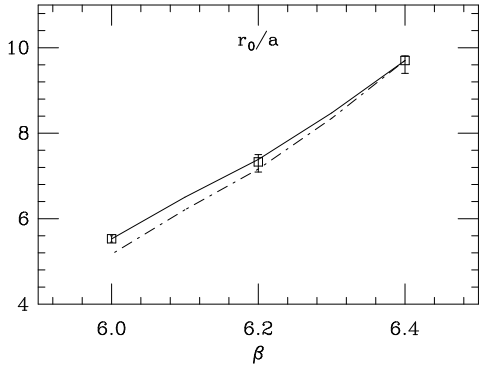

Figure 2. The values for $r_0$ at different $\beta$ compared to the predictions of 2-loop (solid line) and 1-loop (dashed line) perturbation theory.

Table 3
Estimates for $a^{-1}$ [GeV] from $r_0$ and the string tension obtained from the potential ("V") and the Polyakov loop ("PL").

| $\beta$ | $a^{-1}(r_0)$ | $a_V^{-1}$ | $a_{PL}^{-1}$ |
|---|---|---|---|
| 6.0 | $2.23\,^{+3}_{-3}\,^{+0}_{-4}$ | $2.04\,^{+3}_{-3}\,^{+3}_{-3}$ | $2.13\,^{+10}_{-11}$ |
| 6.2 | $2.95\,^{+7}_{-10}\,^{+1}_{-4}$ | $2.77\,^{+3}_{-3}\,^{+3}_{-5}$ | $2.77\,^{+12}_{-13}$ |
| 6.4 | $3.90\,^{+4}_{-12}\,^{+5}_{-1}$ | $3.68\,^{+2}_{-3}\,^{+1}_{-10}$ | $3.50\,^{+11}_{-15}$ |

Our results represent a first step in our attempts to perform a controlled extrapolation to the continuum limit. The analysis of the static quark potential shows a consistent picture for the three values of $\beta$ considered. In particular, it is confirmed that $r_0$ can be obtained reliably from the linear part of the interquark force. These findings need to be corroborated with higher statistics.

I wish to thank S. Booth, D. Henty, M. Lüscher, C. Michael, D. Richards, R. Sommer and M. Teper for useful discussions and suggestions.